\documentclass[twocolumn,pra,showpacs,preprintnumbers,amsmath,amssymb]{revtex4}

\usepackage{graphicx}
\usepackage{dcolumn}
\usepackage{bm}
\begin{document}
\title {Thermodynamical properties of a trapped interacting Bose gas}
\author{Shi-Jie Yang\footnote{Corresponding author: yangshijie@tsinghua.org.cn}}
\author{Yuechan Liu}
\author{Shiping Feng}
\affiliation{Department of Physics, Beijing Normal University,
Beijing 100875, China}
\begin{abstract}
The thermodynamical properties of interacting Bose atoms in a
harmonic potential are studied within the mean-field approximation.
For weak interactions, the quantum statistics is equivalent to an
ideal gas in an effective mean-field potential. The eigenvalue of
the Gross-Pitaevskii equation is identified as the chemical
potential of the ideal gas. The condensation temperature and density
profile of atoms are calculated. It is found that the critical
temperature $T_c$ decreases as the interactions increase. Below the
critical point, the condensation fraction exhibits a universal
relation of $N_0/N=1-(T/T_c)^{\gamma}$, with the index
$\gamma\approx 2.3$ independent of the interaction strength, the
chemical potential, as well as the frequency of the confining
potential.


\end{abstract}
 \maketitle

\section{introduction}
Since the first observations of the Bose-Einstein condensation (BEC)
in dilute alkali-metal atom gases\cite{Anderson,Davis,Bradley},
experimental developments have posed many new tests for many-body
theory. Numerous theoretical approaches have been implemented in
order to obtain accurate results for both the ground-state and
non-equilibrium properties of the trapped boson
systems\cite{Dalfovo,Parkins,Groot,Shi}. Although there is a vast
number of studies in literature, the effects of interactions on the
transition temperature of BEC still remains an open question. The
current consensus among physicists is that the interaction-caused
shift in condensation temperature $\bigtriangleup T_c\equiv
T^{(a)}_c-T^{(0)}_c$, where $T^{(0)}_c$ and $T^{(a)}_c$ are
respectively the critical temperature for ideal and interacting
gases, satisfies a linear relation $\bigtriangleup
T_c/T^{(0)}_c\approx ca_{s} n^{1/3}$ with respect to the $s$-wave
scattering length $a_{s}$. Here $n$ is the particle density.
However, the value and even the sign of the coefficient $c$ is
controversial\cite{Seiringer}. Some authors claimed that repulsive
interactions reduce the critical
temperature\cite{Feynman,Fetter,Toyoda,Giorgini}, while others
indicated an increase of the critical
temperature\cite{Glassgold,Stoof,Holzmann,Huang,Kashurnikov,Arnold,Kastening,Nho,Baym}.
In a recent publication\cite{Betz}, the authors obtained a negative
constant $c\approx-2.33$ by using the model of spatial permutations.
This contradiction is somehow reconciled by a recent
experiment\cite{Smith}, where the researchers demonstrated a
negative shift of the critical temperature for weak interactions and
a positive shift for sufficiently strong interactions.

For an ideal Bose gas in a harmonic trap $V({\bf
r})=\frac{1}{2}m\omega^{2}r^{2}$, the BEC temperature $T_c$ is
determined by $ k_{B}T_{c}=\hbar\omega({N}/{\zeta(3)})^{1/3}$, with
$\zeta(3)=1.202$\cite{Groot,Shi,Minguzzi}. The fraction of
condensation at a temperature $T$ below $T_c$ is $
N_{0}/{N}=1-({T}/T_{c})^{3}$, where $N_0$ is the number of Bose
particles in the condensed state. In this paper, we explore the
phase transition properties of interacting Bose atoms confined in a
spherically symmetric harmonic potential within the mean-field
approximation (MFA). We show that in the weak interaction limit the
particle distribution is equivalent to the ideal gas in an effective
mean-field potential. The eigenvalue of the generalized
Gross-Pitaevskii (GP) equation is identified as the chemical
potential of the ideal gas. For the condensed atoms we obtain the
density profile within the Thomas-Fermi approximation. The
condensation temperature $T_c$ decreases with increase of
interactions as $T_c\sim a_s^{-\beta}$ with $\beta\approx -0.75$, in
contrast to the previous mean-field calculations which gave a linear
relation $\bigtriangleup T_c/T^{(0)}_c\approx -2.47 a_{s}
n^{1/3}$\cite{Giorgini,Smith}. Our results qualitatively coincide
with the recent measurement\cite{Smith}. In particular, the $T_c$ is
independent of confining frequency which implies the details of the
external potential is unimportant to the BEC. Below the critical
temperature, we find a universal relation between the condensed
fraction and the temperature $N_0/N=1-(T/T_c)^{\gamma}$, with an
index $\gamma\approx 2.3$ being independent of the interaction
strength, the chemical potential, as well as the frequency of the
confining potential.

The paper is organized as following: In Sec.II we briefly introduce
the mean-field theory and the related results. Section III displays
our main numerical results. A brief summary is included in Sec.IV.

\section{Mean-field theory}
The Hamiltonian for an interacting Bose gas in a external trap
$V({\bf r})$ is
\begin{equation}
\hat{H}=\int d{\bf
r}\{\hat{\psi}^\dagger[-\frac{\hbar^2}{2m}\triangledown^2+V]\hat\psi+\frac{g}{2}|\hat\psi|^4\},
\end{equation}
where the coupling constant $g=4\pi\hbar^2 a_{s}/m$ is written in
terms of the scattering length $a_{s}$ and atom mass $m$. Following
the standard approach, the Bose field operator is decomposed as
$\hat\psi=\Phi+\hat\phi$, where $\Phi$ is a $c$-number for the
condensate, and $\hat\phi({\bf r})$ is an operator representing its
fluctuations which annihilates a thermal atom at $\bf r$.

Excluding the possibility of aggregate motion and vortices, the
stationary state of $\Phi({\bf r})$ satisfies the generalized GP
equation\cite{Ji,Qi,Liang}
\begin{equation}
-\frac{\hbar^2}{2m}\bigtriangledown^2\Phi+[V({\bf r})-\mu+g(n_0({\bf
r})+2n_T({\bf r}))]\Phi=0,\label{GP}
\end{equation}
where $\mu$ is the eigenvalue of the GP equation. The equation of
motion for $\langle\phi({\bf r},t)\rangle$ is
\begin{equation}
i\hbar\frac{\partial \langle\hat\phi\rangle}{\partial
t}=-\frac{\hbar^2}{2m}\bigtriangledown^2\langle\hat\phi\rangle+[V({\bf
r})-\mu+2gn({\bf r})]\langle\hat\phi\rangle+gn_0({\bf
r})\langle\hat\phi^\dagger\rangle.\label{thermal}
\end{equation}
The condensate density and the number of condensate atoms are
defined by $n_0({\bf r})=|\Phi({\bf r})|^2$ and $N_0=\int d{\bf
r}n_0({\bf r})$, respectively. The noncondensate (thermal) density
is $n_T=\langle\hat\phi^\dagger\hat\phi\rangle$, where
$\langle\cdots\rangle$ indicates a thermal average in the
grand-canonical ensemble. The generalized nonlinear GP equation
(\ref{GP}), which includes interactions between the condensate and
the thermal atoms, is solved for a static condensate, while the
quasiparticle excitations of the system is described by the coupled
Bogoliubov equations (\ref{thermal}), which yields the quasiparticle
energies and amplitudes. These in turn determine the number of
noncondensed atoms $N_T$ as well as various coherence terms
thermodynamic averages over two or more Bose field operators.

As emphasized in Ref.[\onlinecite{Griffin}], the coherence terms
$m_T=\langle\hat\phi\hat\phi\rangle-\Phi^2$ yield an excitation
spectrum that is not gapless. The Popov approximation, which
neglects these terms, reduces the third and fourth-order terms in
$\hat\phi$ and $\hat\phi^\dagger$, respectively, to the first and
second order so that the Hamiltonian can be
diagonalized\cite{Quiroz}. Although it has been successful in
describing the properties of the trapped Bose gases, it is not
well-grounded theoretically, and fails to yield accurate predictions
for the low-lying excitations at high temperatures.

The elementary excitations in Eq.(\ref{thermal}) can be obtained by
employing the semiclassical WKB approximation\cite{Giorgini}. The
coupled equations for $\langle\hat\phi\rangle$ and its complex
conjugate $\langle\hat\phi^\dagger\rangle$, can be solved
explicitly, and the semiclassical excitation spectrum is obtained as
\begin{equation}
\epsilon({\bf p},{\bf r})=\sqrt{[\frac{p^2}{2m}+V({\bf
r})-\mu+2gn({\bf r})]^2-g^2n_0^2({\bf r})}.\label{excitation}
\end{equation}
For a homogeneous system at low temperatures, $n_T$ can be
neglected, and the above excitations coincide with the usual
Bogoliubov spectrum.

The quasiparticles with energies $\epsilon({\bf p},{\bf r})$ are
distributed according to the Bose distribution function,
\begin{equation}
f({\bf p},{\bf r})=\{\exp[\beta\epsilon({\bf p},{\bf r})]-1\}^{-1},
\end{equation}
where $\beta=1/k_B T$. The particle distribution function can be
obtained from the Bogoliubov canonical transformations and is given
by
\begin{equation}
F({\bf p},{\bf
r})=-(\frac{\partial\epsilon}{\partial\mu})_{n_0}f({\bf p},{\bf r}).
\end{equation}

It is notable that the identification of the chemical potential
$\tilde{\mu}$ with the eigenvalue $\mu$ of the GP equation is
questionable in general. In the grand-canonical ensemble, the
chemical potential is defined as $\tilde\mu=\partial E/\partial N$,
corresponding to the energy cost $E$ of adding a particle to the
entire system, not only to the condensate. However, in the framework
of the present theory, the excitation energies (\ref{excitation})
can be approximately reduced to the first order of coupling strength
$g$ as
\begin{equation}
\epsilon({\bf p},{\bf r})\approx\frac{p^2}{2m}+V({\bf
r})-\mu+2gn({\bf r}).
\end{equation}
Therefore, $({\partial\epsilon}/{\partial\mu})_{n}\approx -1$,
yielding the thermal density distribution
\begin{eqnarray}
n_T({\bf r})\approx\int\frac{1}{\exp[\beta(p^2/2m+V_{eff}({\bf
r})-\mu)]-1}\frac{d{\bf p}}{(2\pi\hbar)^3},\label{density}
\end{eqnarray}
This equation is exactly equivalent to the statistics of an ideal
Bose gas confined in an effective mean-field potential,
\begin{equation}
V_{eff}=V({\bf r})+2gn({\bf r}).\label{effective}
\end{equation}
Consequently, the eigenvalue $\mu$ of the GP equation (\ref{GP}) is
justified as the chemical potential of the ideal gas in the
effective potential.

The result can be understood as following: Suppose an interacting
Bose gas is trapped in an external potential $V({\bf r})$ which
reaches an equilibrium distribution $n({\bf r})$. Adding an extra
noninteracting Bose atom increases an energy $\Delta \epsilon({\bf
r})=V({\bf r})+2gn({\bf r})$. As the new comer is assimilated into
the interacting gas, the density distribution $n({\bf r})$ adjust
slightly in order to transfer the increased energy into the
effective potential. The factor $2$ in the effective potential
(\ref{effective}) comes from the Bose symmetry.

\section{numerical results}
At finite temperature $T$ below $T_{c}$, interaction effects
involving the thermal atoms should be taken into account in addition
to those of the condensate, especially when $T$ goes close to
$T_{c}$. Integrating over the momentum ${\bf p}$ in
Eq.(\ref{density}), we obtain in the harmonic potential,
\begin{equation}
n_{T}(\textbf{r})=\sum_{k=1}^{\infty}\frac{e^{-k\beta(m\omega^{2}r^{2}/2+2gn(\textbf{r})-\mu)}}{(\lambda_T\sqrt
k)^{3}},\label{NT}
\end{equation}
where $\lambda_{T}=(2\pi\hbar^{2}/{mk_{B}T})^{1/2}$ is the thermal
de Broglie wavelength.

From the generalized GP equation (\ref{GP}), the Thomas-Fermi
approximation
\begin{equation}
\mu=V(\textbf{r})+g[n_{0}(\textbf{r})+2n_{T}(\textbf{r})],
\end{equation}
will be used in the following discussions. For a spherically
symmetric potential, there is a boundary radius $r_{1}$. For
$r>r_{1}$, the condensate density $n_{0}(\textbf{r})=0$. The thermal
density is determined by,
\begin{eqnarray}
n_{T}(\textbf{r})=\sum_{k=1}^{\infty}\frac{e^{-k\beta[m\omega^{2}r^{2}/2+2gn_{T}(\textbf{r})-\mu]}}{(\lambda_T\sqrt
k)^{3}}.
\end{eqnarray}
Whereas for $r<r_{1}$, $n_{0}(\textbf{r})\neq0$,
\begin{eqnarray}
n_{T}(\textbf{r})=\sum_{k=1}^{\infty}\frac{e^{k\beta[m\omega^{2}r^{2}/2+2gn_{T}(\textbf{r})-\mu]}}{(\lambda_T\sqrt
k)^{3}}.
\end{eqnarray}

\begin{figure}
\begin{center}
\includegraphics*[width=8cm]{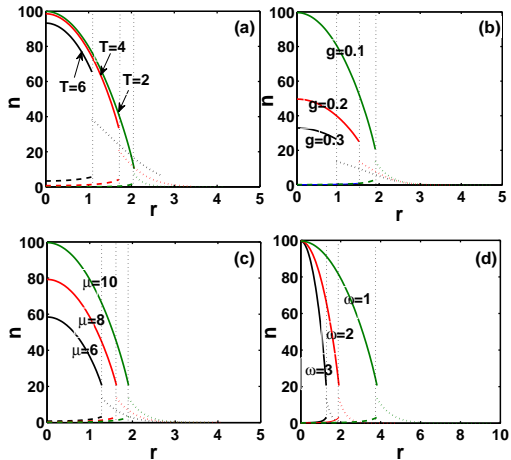}
\caption{(Color online) Density distributions of the condensed (real
curves) and noncondensed (dashed and dotted curves) atoms below the
critical temperature. (a) Temperature dependence at $\mu=10$,
$g=0.1$, and $\omega=2$. (b) Interaction strength dependence at
$\mu=10$, $T=3$, and $\omega=2$. (c) Chemical potential dependence
at $g=0.1$, $T=3$, and $\omega=2$. (d) Potential frequency
dependence at $\mu=10$, $T=3$, and $g=0.1$.}
\end{center}
\end{figure}

In Fig.1, we plot the various cases of the density profiles of
condensed and noncondensed atoms in the spherically symmetric
harmonic potential. Hereafter the units of $\hbar=m=k_B=1$ are used.
For a better understanding of the dependence of the density on
temperature, in Fig.2 we plot the total condensed atoms as a
function of temperature. Figure 2(a) shows that the number of
condensed atoms decreases with increase of interactions at a fixed
chemical potential and a given temperature. Figure 2(b) is the
number of condensed atoms for different chemical potentials at a
given interaction. It coincides with the measurements carried out by
Ensher et al\cite{Ensher}. In particular, as expected, it increases
with the chemical potential.
\begin{figure}
\begin{center}
\includegraphics*[width=8cm]{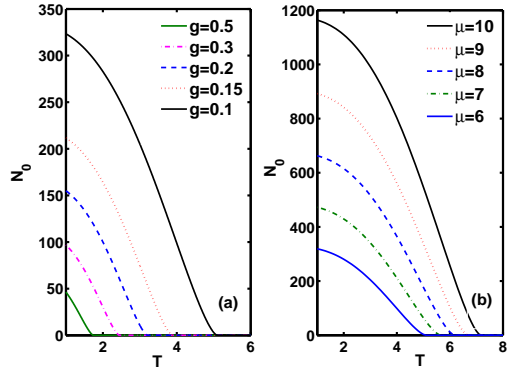}
\caption{(Color online) The condensed atoms versus the temperature
for (a) various interaction strength at $\omega=1$ and $\mu=6$ and
for (b) various chemical potentials at $\omega=1$ and $g=0.5$.}
\end{center}
\end{figure}

\begin{figure}
\begin{center}
\includegraphics*[width=8cm]{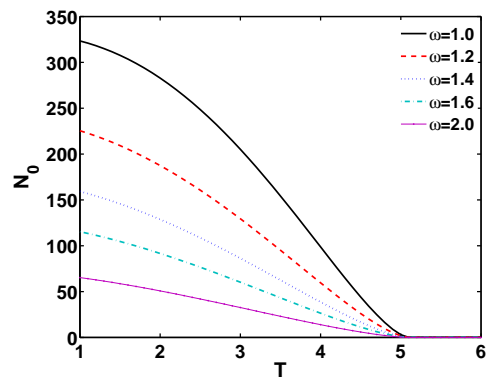}
\caption{(Color online) The condensed atoms versus the temperature
for various frequencies of the confining potential at $g=0.1$ and
$\mu=6$.}
\end{center}
\end{figure}

We note that in our calculations that in our parameter ranges, the
critical temperature is almost independent of the characteristic
frequency $\omega$ of the confining potential. Figure 3 reveals that
all curves meet at one temperature point where the condensed atoms
tend to zero. We consider that this result may be relevant to the
mean-field theory when applying to a finite system, as well as
semi-classical approximation in the calculation of the state
density. Nevertheless, it implies that the condensation temperature
is insensitive to the details of the confining trap.

At the critical temperature $T_{c}$, the condensed atoms is nearly
zero. We obtain the critical temperature from Fig.2 for various
interaction strength and chemical potentials, respectively, as shown
in Fig.4. Figure 4(a) shows that the critical temperature $T_c$
decreases with increase of the interaction strength $g$. The shift
of critical temperature becomes saturated as the interactions
increases. The inset is a log-log plot of the critical temperature
versus the interaction strength. It reveals that the critical
temperature decreases with the interaction strength or $s$-wave
scattering length $a_s$ as $T_c\sim a_s^{\beta}$ with $\beta\approx
-0.75$, in contrast to previous MFA calculations which give a linear
relation $\bigtriangleup T_c/T^{(0)}_c\approx -2.47 a_{s}
n^{1/3}$\cite{Giorgini,Smith}. Our results is qualitatively
consistent with the experimental measurements by Smith et
al\cite{Smith}. Although the mean-field theory has weakness in
dealing with the thermodynamical properties near the critical point,
as discussed in Sec.II, it is remarkable that the results agree with
the experiment quite well. Figure 4(b) shows that the critical
temperature almost linearly increases with the chemical potential
$\mu$.

\begin{figure}
\begin{center}
\includegraphics*[width=8cm]{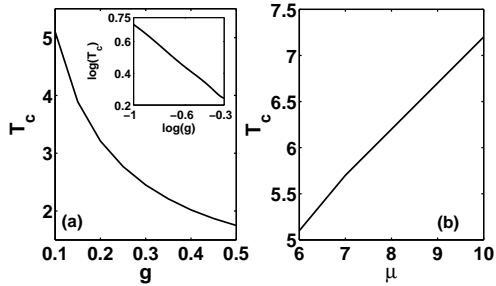}
\caption{(a) The critical temperature versus the interaction
strength for $\omega=1$ and $\mu=6$. Inset: The Log-Log plot gives a
nearly linear relation with a slope $\beta\approx -0.75$. (b) The
critical temperature versus the chemical potential for $\omega=1$
and $g=0.5$.}
\end{center}
\end{figure}

\begin{figure}
\begin{center}
\includegraphics*[width=8cm]{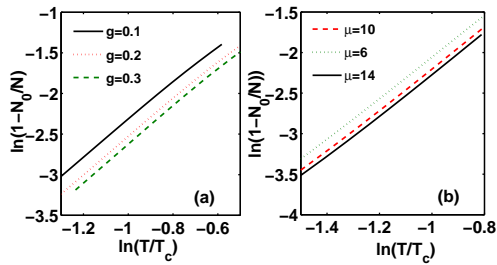}
\caption{(Color online) The log-log plot of the condensate fraction
versus temperature for different interaction strength (a) and for
different chemical potentials (b).}
\end{center}
\end{figure}

To examine the universal behavior below the critical temperature, we
study the temperature dependence of the condensate fraction $N_0/N$
which takes the form,
\begin{eqnarray}
\frac{N_{0}}{N}=1-(\frac{T}{T_{c}})^\gamma.
\end{eqnarray}
Figure 5 is a log-log plot of the condensate fraction versus the
reduced temperature. It shows that for various interaction strength
(Fig.5(a)) and chemical potentials (Fig.5(b)), respectively, the
lines are parallel to each other, giving rise to a common slope
$\gamma\approx 2.3$. This index $\gamma$ is universal which is
independent of the interaction strength, the chemical potential, as
well as the frequency of the confining potential. In comparison, the
ideal gas in a confining potential has an index $\gamma_0=3$, which
is also independent of the characteristic frequency of the confining
potential.

\section{summary}
We have studied the thermodynamical properties of the interacting
Bose gas confined in an external potential. For weak interactions,
the system can be equivalent to an ideal gas in an effective
mean-field potential where the eigenvalue of the generalized GP
equation plays the role of the chemical potential. The critical
temperature decreases nonlinearly with increase of interactions. Our
results agree quite well with the recent experimental measurements.
Furthermore, we found a universal relation of the condensate
fraction with respect to the temperature below the critical
temperature, with the index being independent of the interaction
strength, the chemical potential, as well as the frequency of the
confining potential. We expect verifications of the universal
relation by experiments.

This work is supported by the National Natural Science Foundation of
China under grant Nos. 10874018 and 11074023, the funds from the
Ministry of Science and Technology of China under Grant No.
2012CB821403, and the Fundamental Research Funds for the Central
Universities.

\end{document}